\documentclass[prd,preprint,showpacs,eqsecnum,nofootinbib]{revtex4}
\usepackage{bm}
\usepackage{stmaryrd}
\usepackage{amsmath}
\usepackage{amssymb}
\usepackage{graphicx}
\usepackage{textcomp}
\usepackage{calrsfs}
\usepackage{yfonts}
\usepackage{epsfig}

\newcommand{\nn}{\nonumber}

\newcommand{\be}{\begin{equation}}
\newcommand{\ee}{\end{equation}}
\newcommand{\ben}{\begin{equation*}}
\newcommand{\een}{\end{equation*}}
\newcommand{\bea}{\begin{eqnarray}}
\newcommand{\eea}{\end{eqnarray}}

\newcommand{\bnabla}{\bm{\nabla}}

\begin{document}

\title{{$\mathcal{PT}$-Symmetric Quantum Electrodynamics and Unitarity}}

\date{\today}

\author{Kimball A. Milton}\email{milton@nhn.ou.edu}
\author{E. K. Abalo}\email{abalo@nhn.ou.edu}
\author{Prachi Parashar}\email{prachi@nhn.ou.edu}
\author{Nima Pourtolami}\email{nimap@ou.edu}
\affiliation{Homer L. Dodge Department of
Physics and Astronomy, University of Oklahoma, Norman, OK 73019-2061, USA}
\author{J. Wagner}\email{jeffrey.wagner@ucr.edu}
\affiliation{Department of Physics and Astronomy, University of California, Riverside,
Riverside, CA 92521, USA}

\begin{abstract}
More than 15 years ago, a new approach to quantum mechanics was suggested,
in which Hermiticity of the Hamiltonian was to be replaced by invariance under
a discrete symmetry, the product of parity and time-reversal symmetry, 
$\mathcal{PT}$.  It was shown that if $\mathcal{PT}$ is unbroken, energies
were, in fact, positive, and unitarity was satisifed.  Since quantum mechanics
is quantum field theory in 1 dimension, time, it was natural to extend this
idea to higher-dimensional field theory, and in fact an apparently viable
version of $\mathcal{PT}$-invariant quantum electrodynamics was proposed.
However, it has proved difficult to establish that the unitarity of the
scattering matrix, for example, the K\"all\'en spectral representation for
the photon propagator, can be maintained in this theory.  This has led to
questions of whether, in fact, even quantum mechanical systems are
consistent with probability conservation when Green's functions are
examined, since the latter have to possess physical requirements of analyticity.
The status of $\mathcal{PT}$QED will be reviewed in this report, as well as
the general issue of unitarity.
\end{abstract}

\pacs{03.65.Nk, 12.20.-m, 11.55.Bq, 11.30.Er}
\maketitle

\section{Introduction}
In 1996, following a rather large number of precursors, it began to
be recognized that perhaps the usual requirement of Hermiticity of
the Hamiltonian was overly restrictive \cite{Bender:1996fv}.
Theories described by Lagrangians such as
\be
L=\frac12 \dot x^2-\frac12 x^2-ig x^3 \label{anharmosc}
\ee
were considered, which were suspected of having positive spectra
in spite of the appearance of $i$ in the Lagrangian.  This was
established by numerical calculations shortly thereafter
\cite{Bender:1998ke}, and proved in 2001 \cite{Dorey:2001hi, Dorey:2001uw}.
There remained the question of unitarity, or probability conservation,
and that was established in the following year \cite{Bender:2002vv}.
This required the determination of a new operator $\mathcal{C}$, in
terms of which the inner product was given by taking a new kind of
$\mathcal{CPT}$ inner product.  A method of constructing $\mathcal{C}$
perturbatively was given in Refs.~\cite{Bender:2004vn, Bender:2004sa},
where the first extensions of these ideas to higher-dimensional field
theory were presented.    

Immediately, it was attempted to apply these ideas for a new type of field
theory to the simplest (and, by far, the most successful) gauge theory,
quantum electrodynamics or QED \cite{Bender:1999et}.  It was soon
recognized, however, that this theory would not be renormalizable because
of anomalies (the electric current was an axial vector), but then
another version of $\mathcal{PT}$QED was proposed \cite{milton04} with an ordinary
vector current interaction, but with anomalous transformation properties
under parity.\footnote{This was true parity, not ``intrinsic parity''
suggested in Ref.~\cite{Bender:2003wy}.}  The $\mathcal{C}$ operator,
in lowest order, was constructed the following year \cite{bender05},
which should guarantee the unitarity of the theory.
However, that has proved extraordinarily difficult to establish
\cite{milton11}.

In this paper we review the construction of $\mathcal{PT}$QED in
Sec.~\ref{sec2}.  Then, in Sec.~\ref{sec3} we give the construction
of the $\mathcal{C}$ operator.  In Sec.~\ref{sec4} we give the
leading term in the equivalent Hermitian Hamiltonian, constructed
through this $\mathcal{C}$ operator. This Hamiltonian is then used
to compute, in agreement with the $\mathcal{PT}$ Hamiltonian, the
effective magnetic moment coupling expected from Dirac theory,
but it is noted that this Hermitian Hamiltonian
is extremely cumbersome in practice.  
In Sec.~\ref{sec6} perturbation theory is
used to compute the lowest order vacuum polarization operator,
which fails to exhibit the required K\"all\'en analyticity.
That this seems a general difficulty is exhibited in Sec.~\ref{sec7},
where the $\mathcal{PT}$-symmetric cubic anharmonic oscillator is
considered, and shown to exhibit a failure of unitarity.
The conclusion summarizes the status of our understanding of these 
difficulties.

\section{$\mathcal{PT}$-symmetric quantum electrodynamics}
\label{sec2}
At the first International Workshop on Pseudo-Hermitian Hamiltonians
in Quantum Physics (PHHQP) a $\mathcal{PT}$-symmetric version
of quantum electrodynamics was proposed \cite{milton04}.
A non-Hermitian but $\mathcal{PT}$-symmetric electrodynamics
is based on the assumption of novel transformation properties of the
electromagnetic fields under parity transformations. That is, we assert that
\bea
\mathcal{P}:\quad&&{\bf E(x}, t)\to {\bf E(-x},t),\quad 
{\bf B(x},t)\to- {\bf B(-x},t),\nn\\
&&{\bf A(x},t)\to {\bf A(-x},t),
\quad A^0({\bf x},t)\to-A^0({\bf -x},t),
\label{1}
\eea
which is just the statement that
 the four-vector potential is assumed to transform as an axial
vector.  Under time reversal, the transformations are assumed to be
conventional,
\bea
\mathcal{T}:\quad&& {\bf E(x},t)\to {\bf E(x},-t),\quad {\bf B(x},t)\to- 
{\bf B(x},-t), \nn\\ &&{\bf A(x},t)\to -{\bf A(x},-t),
\quad A^0({\bf x},t)\to A^0({\bf x},-t).
\label{2}
\eea
Fermion fields are assumed to transform conventionally.
This was discussed in detail at the London and Hangzhou PHHQP workshops
\cite{milton11}.

The Lagrangian of the theory then possesses an imaginary coupling
constant in order that it be invariant under the product of these
two symmetries:
\be\mathcal{L}=
-\frac14F^{\mu\nu}F_{\mu\nu}
+\bar\psi\gamma^\mu\frac1i\partial_\mu\psi-m\bar\psi\psi
+ie\bar\psi\gamma^\mu\psi
A_\mu.
\label{3}
\ee

In the radiation (Coulomb) gauge $\bnabla\cdot\mathbf{A}=0$, the dynamical
variables are $\mathbf{A}$ and $\psi$, and the canonical momenta are
$\bm{\pi}_{\mathbf{A}}=-\mathbf{E}^T$, $\pi_\psi=i\bar\psi$.
Here $T$ denotes the transverse part, $\bm{\nabla}\cdot \mathbf{E}^T=0$.
The corresponding Hamiltonian density is
\bea
\mathcal{H}&=&E^2+\mathbf{E}\cdot \bnabla A^0+i\bar\psi\dot\psi
-\mathcal{L}\nn\\
&=&\frac12(E^2+B^2)+\bar\psi\left[\gamma^k\left(
\frac1i\nabla_k+{i}eA_k\right)+m\right]\psi.
\label{4}
\eea

The electric current appearing in both the Lagrangian and Hamiltonian
densities, $j^\mu=\psi^\dagger\gamma^0\gamma^\mu\psi$, transforms
conventionally under both $\mathcal{P}$ and $\mathcal{T}$:
\begin{subequations}
\bea
\mathcal{P}j^\mu({\bf x},t)\mathcal{P}&=&\left(\begin{array}{c}
j^0\\-{\bf j}\end{array}\right)(-{\bf x},t),\label{5a}\\
\mathcal{T}j^\mu({\bf x},t)\mathcal{T}&=&\left(\begin{array}{c}
j^0\\-{\bf j}\end{array}\right)({\bf x},-t).
\label{5b}
\eea
\end{subequations}

Since we are working in the Coulomb gauge, $\bm{\nabla}\cdot{\bf A}=0$, the
nonzero canonical equal-time commutation relations are
\begin{subequations}
\bea
\{\psi_a({\bf x},t),\psi_b^\dagger({\bf y},t)\}&=&\delta_{ab}\delta({\bf x-y}),
\label{6a}\\
{}[A_i^T(\mathbf{x}),E_j^T(\mathbf{y})]&=&-i\left[\delta_{ij}-\frac{\nabla_i
\nabla_j}{\nabla^2}\right]\delta(\mathbf{x-y}).
\label{6b}
\eea
\end{subequations}
We will implicitly assume in the following that the electric field is 
transverse.

\section{The $\mathcal{C}$ operator}\label{sec3}
As for quantum mechanical systems, and for scalar quantum field
theory \cite{Bender:2004vn,Bender:2004sa}, to define a positive norm, 
we seek a $\mathcal{C}$ operator in the form
\be
\mathcal{C}=e^Q \mathcal{P},
\label{7}
\ee
where $\mathcal{P}$ is the parity operator.  $\mathcal{C}$ must
satisfy the properties
\begin{subequations}
\bea
\mathcal{C}^2&=&1,\label{8a}\\
{}[\mathcal{C},\mathcal{PT}]&=&0,\label{8b}\\
{}[\mathcal{C},H]&=&0.\label{8c}
\eea
\end{subequations}
From the first two equations we infer
\begin{subequations}
\be
Q=-\mathcal{P}Q\mathcal{P},\label{9a}\ee
and because $\mathcal{PT}=\mathcal{TP}$,
\be
Q=-\mathcal{T}Q\mathcal{T}.\label{9b}
\ee
\end{subequations}
The third equation (\ref{8c}) allows us to determine $Q$ perturbatively.
If we separate the interaction part of the Hamiltonian from the free part,
\be
H=H_0+eH_1,\label{10}
\ee
and assume a perturbative expansion of $Q$:
\be Q=eQ_1+e^2 Q_2+\dots,\label{11}
\ee
the first contribution to the $Q$ operator is determined by
\be
[Q_1,H_0]=2H_1.\label{12}
\ee
The second correction commutes with the Hamiltonian,
\be
[Q_2,H_0]=0.\label{13}
\ee
Thus we may take
\be
Q=eQ_1+e^3 Q_3+\dots,\label{14}
\ee
which illustrates a virtue of the exponential form.  The $Q_3$ term
is constrained by
\be
\big[Q_3,H_0]=\frac{1}{6}\big[\big[H_1,Q_1\big],Q_1\big].
\ee

For $\mathcal{PT}$ quantum electrodynamics,
the interaction term in the Hamiltonian is simply the standard QED
interaction term multiplied by $i$,
\be
  H_1=i\int d^3x A_\mu(x) \psi^\dagger(x)\gamma^0\gamma^\mu \psi(x).
\ee
The $Q_1$ operator was calculated in Ref.~\cite{bender05}:
($\mathbf{p+q+r=0}$)
\be
  Q_1=\int \frac{d^3p\,d^3q}{(2\pi)^6}
  {\mathbf{E}(-\mathbf{p}) \brace \mathbf{B}(-\mathbf{p})}\cdot
\psi^\dagger(\mathbf{q})
  \bm{\Gamma}_{ E\brace B}(\mathbf{p,t})\psi(-\mathbf{r}),\label{q1}
\ee
where with $\mathbf{t=r-q}$, $\mathbf{k=p\times t}$, and $\Delta
=4m^2p^2+k^2$,
\begin{subequations}
\bea
  \bm{\Gamma}_E(\mathbf{p,t})&=&\frac{2}{\Delta}\Bigg[
    -i\mathbf{k}\gamma_5-2im\gamma_5\bm{\gamma}\times\mathbf{p}
- \frac{\mathbf{(p\times k)}}{p^2}\gamma^0
\bm{\gamma}\cdot \mathbf{t}
+\frac{2m}{p^2}\gamma^0\mathbf{p\times k}\Bigg],\\
  \bm{\Gamma}_B(\mathbf{p,t})&=&\frac{2}{\Delta}\Bigg[
 -2m\bm{\gamma}\times\mathbf{p}+\frac{\mathbf{p\cdot t}}{p^2}\mathbf{k}+
\frac{i}{p^2}\gamma^0\gamma_5\bm{\gamma}\cdot\mathbf{p}\mathbf{p\times
k}\bigg].
\eea
\end{subequations}

\section{The Hermitian Hamiltonian}\label{sec4}
Mostafazadeh has shown that a $\mathcal{PT}$ theory is equivalent to a
Hermitian theory through a similarity transformation
\cite{Mostafazadeh:2001nr, Mostafazadeh:2002hb, Mostafazadeh:2002id, 
Mostafazadeh:2003gz}. 
For the case where the inner product is
constructed with the $\mathcal{C}$ operator,
with the construction (\ref{7}),  the similarity
transformation can be given by
\be
  h=e^{-\frac{Q}{2}}H \,e^{\frac{Q}{2}}.
\ee
Here, $Q$ has a perturbative structure in odd powers of $e$, as
seen in Eq.~(\ref{14}).
 Using the $Q_1$ operator given in Eq.~(\ref{q1}), we can write the equivalent Hermitian Hamiltonian
as
\be
  h=H_0+\frac{e^2}{4}[H_1,Q_1]+\mathcal{O}(e^4).
\ee
Thus, carrying out the commutations, we find
 the leading term in the equivalent Hermitian Hamiltonian is
given by the following nonlocal expression
\bea
h_{\rm int}&=&\frac{e^2}4[H_1,Q_1]
=i\frac{e^2}4\int  \frac{d^3p'd^3q'd^3r'
    d^3p\,d^3q\,d^3r}{(2\pi)^{9}}
    \delta(p'+q'+r')\delta(p+q+r) \nn\\ &&\times\bigg\{ 
    - i \delta(p-p') \psi^\dagger(-q')\gamma^0\gamma^i\psi(r')
    \psi^\dagger(q)\Gamma^i_E(p,t)\psi(-r) \nn\\  
    &&\quad\mbox{}+ \delta(r'-q)E^i(-p)A^m(p')
    \psi^\dagger(-q')\gamma^0\gamma^m\Gamma^i_E(p,t)\psi(-r) \nn\\ 
    &&\quad\mbox{}- \delta(r-q')E^i(-p)A^m(p') \bigg.
    \psi^\dagger(q)\Gamma^i_E(p,t)\gamma^0\gamma^m\psi(r') \nn\\ 
 &&\quad\mbox{}+ \delta(r'-q)B^i(-p)A^m(p') \bigg.
    \psi^\dagger(-q')\gamma^0\gamma^m\Gamma^i_B(p,t)\psi(-r) \nn\\ 
    &&\quad\mbox{}- \delta(r-q')B^i(-p)A^m(p')
    \psi^\dagger(q)\Gamma^i_B(p,t)\gamma^0\gamma^m\psi(r') \bigg\}.\label{hh}
\eea


When one uses the Hermitian Hamiltonian, one also needs to shift the fields:
\be
\mathbf{a}=e^{-Q/2}\mathbf{A}e^{Q/2}\approx\mathbf{A}
-\frac{e}2[Q_1,\mathbf{A}]+\dots,
\ee
which gives rise to additional fermion-loop graphs. 


In fact it is just from the shift in the field that the Dirac
magnetic moment makes its appearance.  In the $\mathcal{PT}$-symmetric
Dirac equation, the latter comes from
\begin{subequations}
\bea
(m-\gamma\cdot \Pi)(m+\gamma\cdot \Pi)&=&\Pi^2+m^2-\frac{ie}2\sigma^{\mu\nu}
F_{\mu\nu},\\
E^2&=&\bm{\Pi}^2+m^2-ie\bm{\sigma}\cdot \mathbf{B}.\label{dirac}
\eea
\end{subequations}
The shift in $\mathbf{A}$ from Eq.~(\ref{q1}) is seen to be
\be
\delta \mathbf{A}(\mathbf{p})=-\frac{ie}2\int(d\mathbf{r})\psi^\dagger(
\mathbf{r})\bm{\Gamma}_E(-\mathbf{p})\psi(\mathbf{r-p}).\ee
This changes the magnetic energy density in $\mathcal{H}_0=\frac12(E^2+B^2)$,
\bea
\delta \mathcal{H}_0(\mathbf{y})&=&-2mie\mathbf{B}\cdot\int (d\mathbf{x})
(d\mathbf{z})\psi^\dagger(\mathbf{x})
\int\frac{(d\mathbf{p})}{(2\pi)^3}e^{i\mathbf{p\cdot(x+y)}}\nn\\
&&\times \int\frac{(d\mathbf{r})}{(2\pi)^3}e^{i\mathbf{r\cdot(x-z)}}
\frac{i\gamma_5\mathbf{p}\times(\bm{\gamma}\times\mathbf{p})}
{4m^2\mathbf{p}^2+(\mathbf{p+r})^2}\psi(\mathbf{z}).
\eea
Using transversality and the static approximation $|\mathbf{r}|\ll m$,
we get the Dirac moment
\be
\delta \mathcal{H}_0\to -i\frac{e}{2m}\psi^\dagger\gamma^0\bm{\sigma}\cdot
\mathbf{B}\psi,
\ee
which has the extra factor of $i$ seen in Eq.~(\ref{dirac}).  This itself suggests a problem
with the reality of the spectrum of the Dirac electron in an external
``magnetic'' field.

 Unfortunately, to calculate the Schwinger correction to $g-2$ we
would have to work out $Q_3$!  It is much harder to do calculations with
the ``Hermitian'' theory.

\section{Vacuum polarization}
\label{sec6}
\begin{figure}
\centering
\epsfig{file=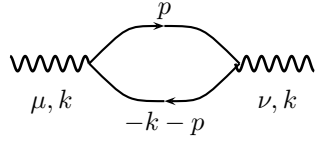}
\caption{Lowest-order vacuum polarization graph}
\label{fig4.loop}
\end{figure}

Let us calculate the correction to the photon propagator  in the 
$\mathcal{PT}$ theory (elsehwere, we will
present calculations in the equivalent Hermitian theory).
In the former, the polarization operator is given by the graph shown
in Fig.~\ref{fig4.loop}.
This corresponds to the amplitude
\be
\int\frac{(dk)}{(2\pi)^4}\frac12A_\mu(-k)A_\nu(k)\Pi^{\mu\nu}(k),
\ee
in terms of the polarization operator
\be
\Pi^{\mu\nu}(k)=e^2\int\frac{(dp)}{(2\pi)^4}\mbox{Tr\,}\frac1{\gamma p+m}
\gamma^\mu\frac1{\gamma(p-k)+m}\gamma^\nu,
\ee
which is opposite in sign to the usual correction.

This leads by any of various standard methods to the following gauge-invariant
form for the corrected renormalized
 photon propagator ($\alpha=e^2/\hbar c$)
\be
\bar D_+(k)=\frac1{k^2-i\epsilon}-\frac{\alpha}{3\pi}
\int_{4m^2}^\infty \frac{dM^2}{M^2}\sqrt{1-\frac{4m^2}{M^2}}\left(1+
\frac{2m^2}{M^2}\right)\frac1{k^2+M^2-i\epsilon}.
\ee
This exhibits a problem with unitarity, because the sign of the imaginary
part is reversed.  That is, on general grounds \cite{Kallen:1972pu,
Schwinger:1974} we should be able to write the full photon propagator
(two-point function) as
\be
\bar D_+(k)=\frac1{k^2}+\int_{4 m^2}^\infty \frac{dM^2}{M^2}
\frac{a(M^2)}{k^2+M^2
-i\epsilon},
\ee
where the spectral function $a(M^2)$ should be positive, since it corresponds,
in lowest order, to the production rate for electon-positron pairs.
In other words,
the generating function for quantum field theory is the vacuum-to-vacuum
persistence amplitude, which reads for the two-point function:
\be
\langle 0_+|0_-\rangle^J=\exp\left[\frac{i}2
\int\frac{(dk)}{(2\pi)^4}J^\mu(-k)\bar D_+
(k)J_\mu(k)\right].
\ee
The probability requirement is \cite{Schwinger:1970xc}
\be
|\langle 0_+|0_+\rangle^J|^2\le1 \Rightarrow 
\mbox{Im\,}\bar D_+(k)\ge0,
\ee
which uses  $k^\mu J_\mu(k)=0$.
This requirement is violated when, as here, $\alpha\to-\alpha$ as compared
to the conventional theory.


\section{Generic Unitarity Problem}
\label{sec7}
This unitarity problem seems to be generic in any quantum theory in 
the $\mathcal{PT}$-framework.
The free harmonic oscillator,
with Lagrangian
\be
L_0=\frac12 \dot x^2-\frac12 x^2,
\ee
 is described by the vacuum persistence amplitude
(generating function)
\be
Z_0[K]=\langle 0_+|0_-\rangle^K=e^{\frac{i}2\int dt\,dt'\,
K(t)\Delta_+(t-t')K(t)},
\ee
in terms of a real source function $K$, with the free causal Green's function
\be
\Delta_+(t-t')=\int\frac{dp}{2\pi}\frac{e^{ip(t-t')}}{-p^2+1-i\epsilon}=
\frac{i}2e^{-i|t-t'|}.\label{prop}
\ee
In terms of the Fourier transform of the source,
\be
\tilde K(p)=\int dt \,e^{-ipt} K(t),
\ee
the generating function is 
\be
\langle 0_+|0_-\rangle^K=\exp\left(\frac{i}2\int\frac{dp}{2\pi}\frac{|\tilde
K(p)|^2}{1-p^2-i\epsilon}\right),
\ee
 so the probability requirement
\be
|\langle 0_+|0_-\rangle^K|\le1\Rightarrow\mbox{Im}\,\Delta_+(p)\ge0,
\ee
is satisfied:
\be
|\langle 0_+|0_-\rangle^K|^2=\exp\left(-\frac12|\tilde K(1)|^2\right).
\ee  The same result is also obtained by using the coordinate-space propagator
(\ref{prop}).

\begin{figure}
\centering
\epsfig{file=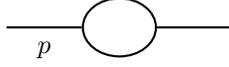}
\caption{Graph contributing to the two-point function in the $igx^3$ theory.}
\label{fig5.loop}
\end{figure}

 However, with a $\mathcal{PT}$-symmetric interaction (\ref{anharmosc}),
the graph shown in Fig.~\ref{fig5.loop} has the wrong sign for the residue of
the pole, as we see by writing the mass operator,
\bea \Sigma(p)=-i\frac{(6ig)^2}{2}\int\frac{dl}{2\pi}\frac1{-l^2+1-i\epsilon}
\frac1{-(l-p)^2+1-i\epsilon}=-\frac{18 g^2}{4-p^2-i\epsilon},\label{fdqm}
\eea
simply evaluating the integral using the residue theorem.
Note that in carrying out perturbative calculations, the $\mathcal{C}$ 
operator,
ostensibly necessary for unitarity, does not make any explicit appearance.
From the mass operator, the corrected propagator is calculated from
\be
\bar\Delta_+(p)=\frac1{-p^2+1-i\epsilon}+\frac1{-p^2+1-i\epsilon}\Sigma(p)
\frac1{-p^2+1-i\epsilon},
\ee
for which the residue of the pole at $p^2=4$ is $2g^2$ ($=18 g^2/9$).
 
In ordinary quantum mechanics, the analog of the K\"all\'en-Lehman spectral
representation for the two-point function \cite{Kallen,Lehmann} is,
in Minkowskian momentum space, for a system having only discrete 
energy states, with energies $E_n$, and eigenvectors $|n\rangle$,
\cite{Bender:2002eg}
\be
\Delta_+(p)=\sum_{n=1}^\infty \frac{Z_n}{M_n^2-p^2-i\epsilon},
\ee
where the spectral masses are
\be M_n=E_n-E_0,
\ee
the energy differences from the ground state, and the residues of the
poles, the ``wavefunction renormalization constants,'' are given
by
\be
Z_n=2M_n|\langle0|x(0)|n\rangle|^2,
\ee which are necessarily non-negative.  The $Z_n$ satisfy a sum rule
\be
\sum_{n=1}^\infty Z_n=1.\label{sr}
\ee

To apply this theorem to $\mathcal{PT}$ theories, we must remove the absolute
value sign in the expression for $Z_n$ (which can also be done in the conventional
theory if real wavefunctions are understood), and write the matrix element
there as an integral which may be extended appropriately into the complex plane:
\be
Z_n=2M_n(-1)^n\left[\int dx\, \psi_0(x) x\psi_n(x)\right]^2,
\ee 
where it is assumed that the wavefunctions are normalized
\be
\int dx\, \psi_0^2(x)=1,\quad \int dx\, \psi_n^2(x)=(-1)^n.
\ee
(It was the indefiniteness of the latter that necessitated the introduction of the
$\mathcal{C}$ operator.)

In Ref.~\cite{Bender:2002eg}, the $\pm gx^4$ theories were examined, and it
was shown that in both cases the sum rule (\ref{sr}) was satisfied. 
Perturbatively the $Z_{2n+1}$ residues were all positive, while the
$Z_{2n}$ terms vanished (by parity), and, for example, the leading
contribution to $Z_3$ is
\be
Z_3=\frac98 g^2+O(g^3).
\ee
This corresponds to the Feynman graph with three internal lines, 
hence a zeroth-order mass $M_3= 3$.
The vanishing of the $Z_{2n}$, however, was not true for larger
$g$ and indeed numerically it was found that the $Z_{2n}$'s were substantial and negative.
See also the Appendix of Ref.~\cite{Bender:2008gh}.

Here we have examined the $igx^3$ theory.  We find, in agreement with the
Feynman diagram calculation (\ref{fdqm}) that 
\be
Z_2=-2g^2+O(g^3).\label{z2}
\ee
The negative sign here, and in Eq.~(\ref{fdqm}), indicates, apparently, a
breakdown of perturbative unitarity.

Let us supply details of the latter calculation, which exhibits some 
interesting subtleties.  The spectral residue for the two-point function
is given by, in Dirac notation
\be 
Z_2=2 M_n \langle0|x|2\rangle\langle 2|x|0\rangle,
\ee
where the states refer to the $\mathcal{PT}$ Hamiltonian
\be
H=H_0+ig H_1, \quad H_0=\frac12 p^2+\frac12 x^2,\quad H_1=x^3.
\ee
It is convenient to transform to the corresponding Hermitian theory
\cite{Bender:2005sc},
\be \tilde H=e^{-Q/2}He^{Q/2},\quad Q=-2g\left(\frac23 p^2+px^2+xpx+x^2p
\right)+O(g^3).
\ee
So the spectral residue is
\be
Z_2 = 2M_2 \tilde{\langle 0|}\tilde x\tilde{|2\rangle}
\tilde{\langle 2|}\tilde x\tilde{|0\rangle},\label{spectralresidue}
\ee
in terms of the Hermitian states and operators, where \cite{Bender:2005sc}
\be \tilde x=e^{-Q/2}xe^{Q/2}=x-ig(x^2+2p^2)+O(g^2).\label{shift}
\ee
Now the Hermitian Hamiltonian, given in Ref.~\cite{Bender:2005sc}, is even
in the Hermitian operators $x$ and $p$, although for the purposes here
we only need to note that
\be
\tilde H=H_0+O(g^2),\ee
so the matrix element $\tilde{\langle 0|}x\tilde{|2\rangle}$ vanishes.  Then
the required matrix elements are
\be
\tilde{\langle 0|}\tilde x\tilde{| 2\rangle}=
-ig\tilde{\langle 0|}(x^2+2p^2)\tilde{|2\rangle}
=\frac{ig}2\tilde{\langle 0|}a^2\tilde{|2\rangle}=
\frac{ig}{\sqrt{2}}=\tilde{\langle 2|}
\tilde x\tilde{|0\rangle}.
\ee
When this is inserted into Eq.~(\ref{spectralresidue}), 
we obtain the result (\ref{z2}).  The same result, of course, can be
obtained in the $\mathcal{PT}$ theory, where the result emerges from
the first-order perturbative corrections to the states.  We have also
redone the Feynman diagram calculation, leading to the result (\ref{fdqm}),
using the Hermitian theory, where the Feynman rules are obtained by using
the shifted operator (\ref{shift}).  The result, obtained either in coordinate
or momentum space, is identical to that found in Eq.~(\ref{fdqm}).  
\section{Conclusions}
Schwinger believed that the Green's functions that defined a
quantum theory are properly defined only in the ``attached Euclidean
space'' \cite{Schwinger:1958rd}.
This defines where the singularities in Green's functions must
lie in order that one can perform a ``Euclidean rotation.''
Such a requirement 
may pose an insuperable barrier to the construction of a consistent
QFT based on a $\mathcal{PT}$-symmetric Lagrangian.

The above calculations were perturbative, and perhaps there are nonperturbative
contributions that save unitarity.  Evidences against this comes from the 
Schwinger model, 2-dimensional massless QED.  
The Schwinger model may be
solved exactly, and exhibits a violation of unitarity, as was discussed
in detail in Ref.~\cite{milton11}.

Thus, it appears that there are problems with unitarity 
not only in
$(\mathcal{PT}\mbox{QED})_{2,4}$, but in any quantum mechanical
system.
Analytic properties required by the probability considerations
and the Euclidean postulate seem to be generically violated.
In 1, 2, and 4 dimensions the analyticity requirements 
(for example, the K\"all\'en spectral
representation for the 2-point function) appear to be violated.
And perturbation theory,
where the $\mathcal{C}$ fails to make any appearance \cite{Bender:2005sc},
 evidently does not  give a positive spectrum
to the massless $\mathcal{PT}$-symmetric electrodynamics in two dimensions.
We are trying to clarify the issue by comparison with the equivalent
Hermitian Hamiltonian.  There, too, positivity of the spectral function
is not assured, because of the necessity of shifting the field in passing
to that alternative description.

\acknowledgments
We thank  the US National Science Foundation, the US Department of Energy, and
the Julian Schwinger Foundation
for partial support of this research.  
We also thank P. Mannheim, K. V. Shajesh, 
and M. Schaden for useful discussions.

\end{document}